\documentclass[twocolumn,aps,floatfix]{revtex4}
\usepackage{longtable}
\usepackage{amsmath,bm}%
\usepackage{graphicx}%
\usepackage{epsf,epsfig,latexsym}
\begin{document}
\title{Isotopic equilibrium constants for very low-density and low-temperature nuclear matter}  
\author{J. B. Natowitz$^1$}
\author{H. Pais$^2$}
\author{G. R\"opke$^3$}
\author{J. Gauthier$^1$}
\author{K. Hagel$^1$}
\author{M. Barbui$^1$}
\author{R. Wada$^1$}
\affiliation{$^1$Cyclotron Institute, Texas A$\&$M University, College Station, 
Texas 77843}
\affiliation{$^2$Department of Physics, University of Coimbra, 3004-516 Coimbra, 
Portugal}
\affiliation{$^3$University of Rostock, FB Physik, 18059 Rostock, Germany }

\date{\today}

\begin{abstract}
Yields of equatorially emitted light isotopes, $1\le Z\le 14$, observed in 
ternary fission in the reaction $^{241}$Pu($n_{\rm th}$,f) are employed to 
determine apparent chemical equilibrium constants for  low-temperature and
low-density nuclear matter.  The degree of equilibration and role of medium 
modifications are probed through a comparison of experimentally derived 
reaction quotients with equilibrium constants calculated using a relativistic 
mean-field model employing a universal medium modification correction for the 
attractive $\sigma$ meson coupling.  The results of these comparisons indicate 
that equilibrium is achieved for the lighter ternary fission isotopes.  For 
the heavier isotopes experimental reaction quotients are well below calculated 
equilibrium constants. This is attributed to a dynamical limitation reflecting 
insufficient time for full equilibrium to develop.  The role of medium 
effects leading to yield reductions is discussed as is the apparent 
enhancement of yields for $^8$He and other very neutron rich exotic nuclei. 
\end{abstract}



\maketitle

\section{Introduction}

A  high quality nuclear equation of state (EOS) applicable over a wide 
range of density and temperature is an essential ingredient for reliable 
simulations of stellar matter and astrophysical phenomena.  In recent decades 
many nuclear theory efforts have been devoted to developing such equations and 
many are available in the literature, see Refs.~\cite{oertel17,lattimer91,
shen98,horowitz06,typel10,hempel10,shen11,shen11_1,voskresenskaya12,roepke13,
furusawa13,fsu,pais18,pais19} 
and references therein. The validation of these equations of state usually 
rests on careful comparisons between the results of theoretical simulations 
and astrophysical observations.

At the same time, laboratory studies of nuclear matter at different densities, 
temperatures and isospin content offer some unique possibilities to address 
specific aspects of the nuclear equation of state. Exploiting a variety of 
projectile energies, projectile-target combinations and reaction mechanisms, 
nuclear experimentalists have probed cluster formation and the composition of 
nuclear matter at different densities, caloric curves and phase transitions, 
the density dependence of the symmetry energy and medium effects on nuclear 
binding energies, see Refs.~\cite{pais18,pais19,fsu,chomaz14,
bali14,elliot12,qin12,hempel15} and references therein. 

While isotope mass fractions are commonly used to present the results of EOS 
composition calculations, references~\cite{pais18,pais19,qin12,hempel15}
employed chemical equilibrium constants for production of $Z = 1$ (H) and 
$Z=2$ (He) derived from the experimental isotope yields.  These are more 
robust quantities for testing different equations of state since, at least 
in the low-density ideal limit, they are less dependent upon the choice of 
isotopes included in the EOS model calculations and upon the source asymmetry. 

The thermodynamic reaction quotient $Q$ for the formation of an isotope $^AZ$ 
with  mass number $A$, atomic number $Z$, and neutron number $N=A-Z$,
%
%
is defined such that, 
                         
\begin{equation}
Q =\frac{ \{^AZ\}}{\{p\}^Z\{n\}^N}
\end{equation}
where curly brackets denote the fugacities of the chemical species, i.e. 
the isotope $^AZ$ as well as the protons ($p$) and neutrons ($n$).  Fugacity 
depends on temperature, pressure and composition of the mixture, among other 
things.  The fomulation in terms of fugacities  arises because components 
in non-ideal systems interact with each other. In nuclear EOS models these 
interactions are modeled in a variety of ways~\cite{oertel17,lattimer91,
horowitz06,voskresenskaya12,hempel10,shen11,shen11_1,furusawa13,roepke13,
pais18,pais19,hempel15}. The right-hand side of this equation corresponds 
to the reaction quotient for arbitrary values of the fugacities.  The reaction 
quotient becomes the equilibrium constant, $K$, if the system reaches 
equilibrium.  The equilibrium constant is related to the standard Gibbs free 
energy change for the reaction, $\Delta G^0$, as 
\begin{equation}
\Delta G^0  =-RT\ln{K}
\end{equation}
where $T$ is the temperature and $R$ is the gas constant. 

If deviations from ideal behavior are neglected, the fugacities may be 
replaced by concentrations or densities. Employing square brackets to 
indicate concentrations or densities at equilibrium we can designate this 
ratio as chemical constant $K_c$, 
\begin{equation}
K_c= \frac{ [^AZ]}{[p]^Z[n]^N}.
\end{equation}
$K_c$ is defined in an equivalent way to the thermodynamic equilibrium 
constant but with concentrations or densities of reactants and products, 
denoted by square brackets, instead of fugacities. 

The experimental equilibrium constants reported in references~\cite{qin12,
hempel15} demonstrated clearly that, even at densities in 
the 0.003  to 0.03 nucleons/fm$^3$ range, interactions are important and 
experimental equilibrium constant data may be employed to evaluate the 
various theoretical models.    

\section{Analysis of Ternary Fission Yields}

In this paper we report extensions of the measurements of isotopic equilibrium 
constants to a broader range of isotopes at even lower temperature and 
densities.  Specifically we derive isotopic equilibrium constants for 
isotopes produced in ternary fission processes which occur in approximately 
0.3 \% of decays during the spontaneous or thermal neutron induced fission of 
a heavy nucleus~\cite{wagemans91,mills95,iaea00,halpern71,theobald89,mehta73,
serot98,heeg89,koester00,koester99,lestone05,tsekhanovich,shafer95,
rubchenya88,rubchenya82,kopatch02,vorobyev18,goennenwein05,vermote10,
piasecki75,piasecki79,sharma83,ramayya07,schubert92,wagemans08,wuenschel14}. 
Such ternary fission is characterized by emission of an energetic light 
particle or fragment in a direction perpendicular to the axis defined by the 
separating massive fragments, signaling their origin in the region between 
the two nascent heavy fragments at or near the time of scission.  Collectively,
such isotopes, are typically identified in the ternary fission literature as 
''scission'' or ''equatorially'' emitted particles. 

This well identified isolated mechanism facilitates exploration of yields with 
minimal perturbations from collision dynamics. This allows an experimental 
test of the chemical equilibrium hypothesis. If that hypothesis is supported, 
derived equilibrium constants provide information against which various 
proposed equations of state may be tested in the low-density limit.  In this 
regard they constitute the experimental counterpart of theoretical virial 
equations of state which serve as a low-density theoretical baseline for EOS 
calculations~\cite{horowitz06,roepke13}.  Data of sufficient accuracy would 
allow a careful evaluation of the density dependence of fragment-fragment 
interactions and in medium modifications of cluster properties.  See 
reference~\cite{roepke20} for a recent discussion of such effects.

The experimental results of Koester \textit{et al.}, obtained with an on-line 
mass spectrometer, provide a comprehensive data set for ternary fission 
yields for 42 isotopes determined in the reaction 
$^{241}$Pu($n_{\rm th}$,f)~\cite{koester00,koester99}.  In addition, 17 upper 
limits are also reported for yields of other isotopes.  In 
reference~\cite{wuenschel14}, these yields were compared to results of 
calculations made using a model which assumes a nucleation-time-moderated 
chemical equilibrium~\cite{demo97,schmelzer97,wilemski95} in the low-density 
matter which constitutes the neck region of the scissioning system.  
Nucleation approaches have much in common with thermal coalescence 
approaches previously applied to clustering in low-density nuclear 
systems~\cite{mekjian78,hagel00} but explicitly incorporate consideration of 
cluster formation rates. Coalescence of nucleons into clusters is a 
dynamic process requiring time, while the fissioning system exists for a 
limited time span.  A reasonably good fit to the $^{241}$Pu($n_{\rm th}$,f) 
experimental data from references~\cite{koester00,koester99} was obtained 
with the following parameters: Temperature, $T= 1.4$ MeV, density, 
$\rho= 4\times10^{-4} {\rm fm}^{-3}$, proton fraction, $Y_p= 0.34$, nucleation 
time $t_{\rm nuc}= 6400$ fm/$c$ and critical cluster mass, $A_{\rm cr} = 5.4$. 
We note that various previous attempts to evaluate the temperatures 
appropriate to thermal neutron induced ternary fission have led to 
temperatures in the range of 1.0 to 1.4 MeV~\cite{andronenko01,popkiewicz96}. 
For the $^{242}$Pu compound nucleus the proton fraction, $Y_p$, is 0.388. 
The derived value of 0.34 indicates that the region between the separating 
fragments, which dominates the production of the ternary particles, is 
neutron enriched~\cite{wuenschel14,sobotka97}.

\section{Equilibrium constant for $^4{\rm He}$}  
Since all yields in the Koester $^{241}$Pu($n_{\rm th}$,f) data are referenced 
to the yield of $^4$He particles we began by establishing 
the correspondent equilibrium constant for this 
particle.  Determining this equilibrium constant requires accurate yields of 
the neutrons, protons and $^4$He ejected at the time of 
scission.  The equatorial emission origin of these particles must be well 
defined and contributions from other sources (e.g. pre-scission emission, 
polar emission, secondary particle emission) to the total yields be carefully 
removed.  Establishing the yields of equatorial emission requires careful 
exploration of the particle angular distributions relative to the scission 
axis. These measurements are difficult, particularly for the neutrons because 
subsequent evaporation from the fission fragments dominates the neutron yield. 
Fortunately, very precise measurements of these yields have been made by a 
number of extremely competent experimental groups and absolute yields for 
many fissioning isotopes are, in fact, available and tabulated in the 
literature~\cite{wagemans91,mills95,iaea00,halpern71,theobald89,mehta73,
serot98,heeg89,koester00,koester99,lestone05,tsekhanovich,shafer95,
rubchenya88,rubchenya82,kopatch02,vorobyev18,goennenwein05,vermote10,
piasecki75,piasecki79,sharma83,ramayya07,schubert92,wagemans08}.  The 
systematics of ternary fission yields have been extensively analyzed in 
various evaluations and review 
articles~\cite{wagemans91,mills95,iaea00,serot98}.  Focusing particularly on 
values reported for Pu isotopes we have adopted for our calculations the 
experimental values indicated in column 4 of Table \ref{tab:1}.  The adopted 
value for the $^4$He particle yield includes a 17 \% correction to remove 
$^4$He particles resulting from the decay of $^5$He nuclei emitted at 
scission~\cite{goennenwein05}.  The adopted value for protons includes a 
14.5 \% correction to remove polar emission 
protons~\cite{piasecki75,piasecki79,sharma83}.  For neutrons, the adopted 
value is that determined for scission neutrons~\cite{vorobyev18}. 


\begin{table*}[t]
\begin{tabular}{|c|c|c|c|}
\hline
particle & total yield/fission  & equatorial scission emission &  adopted yield \\
\hline
$n$      & $2.96\pm 0.005$        &$0.107\pm 0.015$       &$0.107 \pm 0.015$  \\
$p$      & $4.08\times10^{-5}\pm0.41$ &$3.49\times10^{-5}\pm 0.35$ &$3.49\times10^{-5}\pm 0.35$    \\
$^4$He & $2.015\times10^{-3}\pm 0.20$ &$2.00\times 10^{-3}\pm0.20$&$1.66\times10^{-3}\pm 0.17$ \\
\hline

\end{tabular}
\caption{Adopted values of neutron, proton and $^4$He  
yields~\cite{mills95,iaea00,kopatch02,vorobyev18,goennenwein05,vermote10,
piasecki75,piasecki79,sharma83,ramayya07,schubert92,wagemans08}.  
References are the primary sources. Measurements and 
systematics of other data for adjacent isotopes were also employed in 
establishing these values. Uncertainties are 1$\sigma$.}
\label{tab:1}
\end{table*}

Applying the thermal coalescence model of Mekjian~\cite{mekjian78} to these 
data allows extraction of the coalescence volume, 2937 fm$^3$.  With the 
absolute yields and this coalescence volume, the relevant densities and the 
experimental equilibrium constant $K_c$($^4$He) for direct formation of the 
$^4$He in its ground state is $3.02 (\pm1.07) \times 10^{18}$ fm$^9$.  As 
indicated above, the largest contributor to the uncertainty is the neutron 
scission yield. Note, however, that the apparent effective $K_c^{\rm eff}$ 
for the total experimentally observed $^4$He yield (column 2, 
Table~\ref{tab:1}), which includes the $^5$He contribution (as well as 
possible smaller contributions from other particle unstable isotopes) 
is $3.66(\pm1.30) \times 10^{18}$  fm$^9$.  By convention, relative yields 
in ternary fission are typically normalized to the total $^4$He yield.  

In reference~\cite{pais18} Pais \textit{et al.} reported a study of in-medium 
modifications on light cluster properties, within the relativistic mean-field 
approximation, where explicit binding energy shifts and a modification on 
the scalar cluster-meson coupling were introduced in order to take these 
medium effects into account.  The interactions of the clusters with the 
surrounding medium are described with a phenomenological modification, 
$x_{i,\sigma}$, of the coupling constant to the $\sigma$ meson, 
$g_{i,\sigma}=x_{i,\sigma} A_i g_{\sigma}$.  Using the FSU Gold EOS~\cite{fsu}
and requiring that the cluster fractions exhibit the correct behavior in the 
low-density virial limit ~\cite{horowitz06,voskresenskaya12,roepke13}, they 
obtained a universal scalar cluster-meson coupling fraction, 
$x_{i,\sigma}=0.85\pm0.05$, which could reproduce both  this limit and the 
equilibrium constants extracted from reaction ion data~\cite{qin12,hempel15} 
reasonably well.  The results are qualitatively similar to the ones obtained 
with other approaches~\cite{horowitz06,voskresenskaya12,hempel10,shen11,
shen11_1,hempel15}.  Employing the model of reference~\cite{pais18} with  
$T=1.4$ MeV, $\rho_{\rm tot} = 4 \times 10^{-4}$ fm$^{-3}$, and a scalar 
cluster-meson coupling fraction $x_{i,\sigma}= 0.85$ leads to 
$K_c$($^4$He)= $2.99 \times 10^{18}$ fm$^9$ for direct production, and 
$K_c^{\rm eff}$($^4$He) = $3.65 \times 10^{18}$ fm$^9$. 

In a more recent work ~\cite{pais19}, Pais \textit{et al.} compared their 
model results to equilibrium constants calculated from a new analysis, where 
in-medium modifications are addressed, for experimental data measured in 
intermediate energy Xe + Sn collisions.  This comparison lead to a higher 
scalar cluster-meson coupling constant $x_{i,\sigma}= 0.92\pm0.02$. 

With this higher assumed value of the coupling constant, the in-medium effects 
are reduced, and the predicted value for $K_c$($^4$He) becomes 
$3.75 \times10^{18}$ fm$^9$, and for $K_c^{\rm eff}$ ($^4$He) = 
$4.62 \times 10^{18}$ fm$^9$.

\section{Extension to Other Isotopes}

Using the adopted values of the equatorial neutron and proton yields together 
with the measured yields for all isotopes we have calculated the effective 
experimental reaction quotients, $Q_c^{\rm eff}$ for formation of the observed 
isotopes from the nucleons,
 i. e., 
                                                                         
\begin{equation}
Q_c^{\rm eff} = \frac{ [^AZ]}{[p]^Z[n]^N}
\end{equation}
where eff denotes  total observed yields including all contributions from 
gamma decaying and particle decaying excited states.  Here we employ $Q$ 
because in our previous treatment of these same data within the framework 
of a nucleation time modulated statistical equilibrium model we have presented 
evidence that statistical equilibrium is not achieved for the heaviest 
isotopes~\cite{wuenschel14}.  The term effective is used in recognition of 
the fact that the final observed ground state yields include contributions 
from de-excitation of  short lived gamma or particle decaying states 
initially present in the primary isotope distribution.  The relative 
importance of such contributions will vary with temperature and density.  For 
a system at equilibrium $Q_c^{\rm eff}$ = $K_c^{\rm eff}$, the effective 
equilibrium constant.  A direct comparison between the experimental  results 
and those of theoretical calculations requires that the contributions from 
relevant excited states be included in the theoretical treatment. 

In the original formulation by Pais \textit{et al.}, only ground states 
including particle unstable ground states  were included in the calculation. 
For the present calculation we have included experimentally identified 
(excitation energy and spin) gamma decaying excited states~\cite{ensdf} which 
can have a significant population at $T = 1.4$ MeV.  We have also included 
relevant particle unstable isotopes and states which can feed the observed 
population~\cite{ensdf}.  With this ensemble of states we performed some 
preliminary calculations to explore the sensitivity of various results to the 
assumed density. We found the final free neutron to free proton ratio to be 
very sensitive to density. A comparison of the theoretical free $n/p$ ratio 
to the experimentally observed free $n/p$ ratio for different assumed total 
densities indicated a density of $2.56(\pm0.20) \times 10^{-4} {\rm fm}^{-3}$. 
This value, which is somewhat lower than the $4 \times 10^{-4} {\rm fm}^{-3}$ 
derived from nucleation model fits, has been adopted for the present 
calculations.  In the recent treatment of the emisssion of $Z=1,2$ isotopes 
in the spontaneous ternary fission of $^{252}$Cf a different approach 
suggests quite similar values~\cite{roepke20}.


\begin{figure}
\epsfig{file=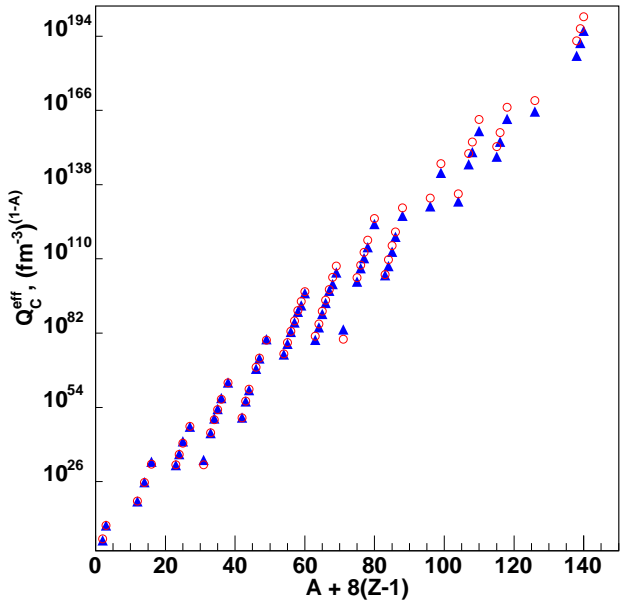,width=9.2cm,angle=0}
 \caption{$Q_c^{\rm eff}$ values vs $A+8(Z-1)$. Triangles - experimental 
results.  Open circles - theoretical results for $K_c^{\rm eff}$ with 
$T=1.4$ MeV, $Y_p = 0.34$, $\rho = 2.56\times10^{-4} {\rm fm}^{-3}$ and 
coupling constant $x_{i,\sigma}= 0.92$.}
\label{fig1}
\end{figure}

The experimentally derived reaction quotients are presented in 
Figure~\ref{fig1}.  To more clearly present the data, we plot $Q_c^{\rm eff}$ 
against the isotope identifier parameter proposed by Lestone~\cite{lestone05}, 
i.e., $A+8(Z-1)$.  For comparison to the experimental $Q_c^{\rm eff}$ values, 
we also present theoretically calculated equilibrium constants, 
$K_c^{\rm eff}$, obtained using the model of Pais \textit{et al.}~\cite{pais18} 
with a scalar cluster-meson coupling constant $x_{i,\sigma}$ of 0.92.  To 
carry out these calculations we fixed the temperature to be 1.4 MeV, the total 
density to be $2.56 \times 10^{-4}$ nucleons/fm$^3$ and the proton fraction 
of the matter to be 0.34. Both the experimental and theoretical values 
are tabulated in Appendix A of this paper.  Unlike the data employed for the 
previous comparisons with this model, the present data  include isotopes as 
heavy as $^{36}$Si.  Therefore the role of excited states should be much more 
important in determining the observed isotope yields.  This is particularly 
true for nuclei with lower energy gamma decaying excited states with high 
degeneracies.  Particle decaying excited states are also included but many 
generally occur at relatively higher excitation energies and thus are less 
populated at low temperature. 

As is observed in Figure \ref{fig1}, the experimental and theoretical trends 
are quite similar. For the heaviest isotopes there is, however a clear 
indication that the experimental $Q_c^{\rm eff}$ values fall well below the 
theoretically calculated equilibrium constants.  To better appreciate these 
differences we plot, in Figure 2 the ratios of the values of the 
experimentally derived reaction coefficients to the $K_c^{\rm eff}$ values 
calculated theoretically using the Pais \textit{et al.} 
formulation~\cite{pais18}.  Ratios for isotopes for which measured 
experimental yield values exist are identified by triangles.  Those for which 
only upper limits to the experimental yields are available are not included 
in this figure.  

\begin{figure}
\epsfig{file=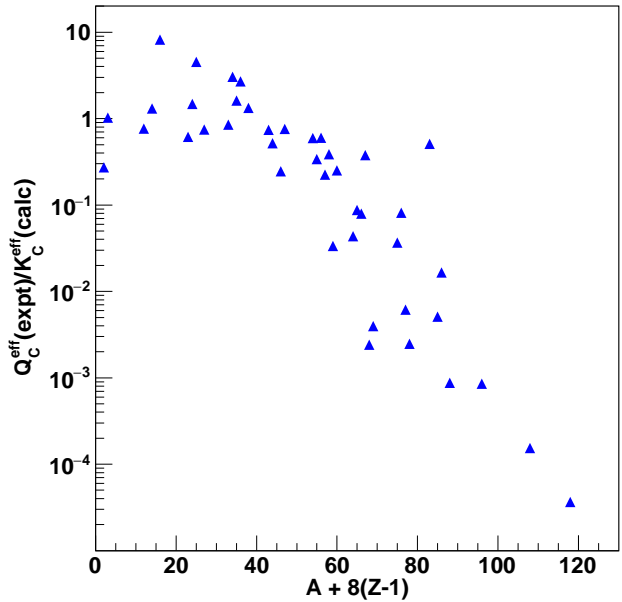,width=9.2cm,angle=0}
 \caption{ Ratio $Q_c^{\rm eff}$ (experiment)/$Q_c^{\rm eff}$ (theory) vs  
$A+8(Z-1)$. All isotopes in Koester data table are considered.  Isotopes for 
which only upper limits are reported are excluded from this plot. See text. }
\label{fig2}
\end{figure}

In Figure \ref{fig2} we see that for the lighter isotopes there is some 
scatter about the ratio $R_{\rm exp/theo} = Q_c^{\rm exp}/Q_c^{\rm theo}=1$, 
but a general overall accord between the data and the theoretical values, 
suggesting that chemical equilibrium has been achieved for the isotopes with 
$Z \leq 5$.  The experimental $K_c$ value reported for the $^2$H is well 
below the theoretical value.  This appears to reflect the very weak binding of 
the deuteron. Such reductions in deuteron yield are a general feature in the 
production of deuterons in heavy ion 
collisions~\cite{pais18,qin12,cervesato92}.  Interestingly, for the light 
neutron rich isotopes $^8$He, $^9$Li, $^{10}$Be and $^{12}$Be the ratio 
indicates significant excesses relative to the calculated values in the 
region where there is a reasonable agreement for the other isotopes.
 
Above that point the plotted ratios drop rapidly falling to 
$R_{\rm exp/theo}\sim10^{-5}$ for the heaviest isotopes. Since the Pais 
calculation includes medium effects through the cluster coupling constant 
this decrease does not appear to reflect calculated medium effects.  Rather, 
the observed decline in the ratio of experimental value to theoretical value 
indicates that equilibrium is not reached for the heavier isotopes.  This is 
entirely consistent with the conclusion reached in reference~\cite{wuenschel14}
where it is attributed to a time moderated nucleation effect.  The possibility 
that finite size effects may also contribute to this decline is not ruled out. 

\section{$Z = 1$ and 2 isotopes and medium modifications}

Given the recent detailed analysis of $Z=1$ (H) and 2 (He) isotope production 
for $^{252}$Cf ternary fission~\cite{roepke20} it is interesting to focus 
explicitly on these results for the present case.  In Table~\ref{Tab:2}, the 
available measured equilibrium constants for these isotopes are presented and 
compared to the theoretical values calculated using a scaler cluster-meson 
coupling constant $x_{i \sigma}= 0.92$. As already noted above, the 
experimental $Q_c$ value based on the observed yield for the $^2$H is well 
below the theoretical value.  (This is also true in the $^{252}$Cf 
case~\cite{roepke20}.)  This suggests a clear medium effect for this very 
weakly bound nucleus~\cite{roepke82,hagel12}. $^3$He was not observed in 
the Koester experiment nor has a $^3$He yield been reported in any other 
ternary fission experiment~\cite{goennenwein05,vermote10}.  The theoretically 
calculated  $^3$He and $^3$H equilibrium constants in Table~\ref{Tab:2} are  
similar, as expected, the difference arising from the small binding energy 
difference for these $A=3$ isotopes.  While some similar medium effect may 
operate on the $A=3$ yields, the non-observation of $^3$He reflects the very 
small free proton to free neutron ratio at equilibrium indicated in Table 1. 
Given that ratio, the $^3$He yield should be about four orders of magnitude 
below the $^3$H yield.  This low yield, together with possible additional 
factors specific to individual experiments. e.g., separation, identification 
and background discrimination, offers a natural explanation for the absence 
of $^3$He yield data in the literature. 

\begin{table}[t]
\begin{tabular}{|c|c|c|}
\hline
particle & $Q_c^{\rm eff}$ (expt)           &$K_c^{\rm eff}$ (calc)         \\
\hline
$^{2}$H  & $5.50(\pm0.99)\times10^{3}$  & $2.42\times10^{4}$  \\
$^{3}$H  & $2.84(\pm0.85)\times10^{9}$  & $3.29\times10^{9}$  \\
$^{3}$He &                              & $1.43\times10^{9}$  \\
$^{4}$He & $3.66(\pm1.30)\times10^{18}$ & $4.74\times10^{18}$ \\
$^{5}$He &                              & $1.50\times10^{22}$ \\
$^{6}$He & $5.95(\pm3.50)\times10^{25}$ & $5.45\times10^{25}$ \\
$^{7}$He &                              & $2.60\times10^{28}$ \\
$^{8}$He & $2.60(\pm1.97)\times10^{33}$ & $3.76\times10^{32}$ \\
\hline
\end{tabular}
\caption{Chemical constants for the isotopes of the light elements H, He. The experimental values $Q_c^{\rm eff}$ (expt)
are compared to calculated values $K_c^{\rm eff}$ (calc).}
\label{Tab:2}
\end{table}

For $^3$H, $^4$He and $^6$He the tabulation indicates reasonable agreement 
(within experimental errors) between experiment and theory. 

In contrast the experimental value for the very neutron rich $^8$He is an order
of magnitude higher than that calculated. The large experimental yield of 
$^8$He is a general feature of ternary fission experiments.  This special 
nature of $^8$He may reflect some feature of dynamics, e.g., time dependent 
density or temperature fluctuations or feeding from parent nuclei, or of 
detailed structural features not yet understood.  As noted in the previous 
section the comparison of the experimental equilibrium constants with those 
of the calculation (Figures \ref{fig1}, \ref{fig2}, Tables \ref{Tab:A1}, 
\ref{Tab:A2}) also indicates yield enhancements for the other neutron rich 
isotopes $^9$Li, $^{10}$Be and $^{12}$Be.  The cluster structure of such 
neutron rich nuclei has been discussed in the framework of an extended Ikeda 
diagram~\cite{oertzen11}.  Particularly intriguing is the possibility that 
the yield enhancement reflects the existence of strong neutron correlations 
in the disassembling matter.  In this regard $^8$He is of special interest 
as experimental evidence for a possible alpha-tetra-neutron structure has 
been published~\cite{kisamori16} and some theoretical work suggests 
that a tetra-neutron condensate might be formed in low-density neutron rich 
stellar matter~\cite{ivanytskyi19}.  This subject warrants further 
investigation. 

In a recent related paper on the spontaneous ternary fission of 
$^{252}$Cf~\cite{roepke20}, we explored an alternative information entropy 
based analysis to characterize the emission of $Z = 1, 2$ isotopes as a 
basis for evaluating medium effects.  In that paper it was proposed that 
relevant primary distribution of isotopes formed in the ternary fission 
process could be characterized by a few Lagrange parameters $\lambda_T$, 
$\lambda_n$ , $\lambda_p$  such that
\begin{equation}
Y_{A,Z}^{\rm rel}\propto R_{A,Z}^{\rm rel}\,\, g_{A,Z}\left(\frac{2\pi\hbar^2}{Am\lambda_T}\right)^{-\frac{3}{2}}
\!\!e^{[B_{A,Z} + (A-Z)\lambda_n + Z\lambda_p]/\lambda_T}
\label{Yrel}
\end{equation}
where $g_{A,Z}$ denotes the degeneracy of the nucleus $\{A,Z\}$ in the ground 
state, $B_{A,Z}$ its binding energy, $m$ is the average nucleon mass.  The 
Lagrange parameters $\lambda_i$ are non-equilibrium generalizations of the 
equilibrium thermodynamic parameters $T$, $\mu_n$, $\mu_p$.   Different 
approximations to treat the Hamiltonian of the many-nucleon system lead to 
different values for these parameters. In particular all relevant excited 
states and continuum states have to be taken into account, and in-medium 
mean-field and Pauli blocking effects must be included.  These effects are 
collected in a prefactor $R^{\rm rel}_{A,Z}$ which, in general, depends on 
the Lagrange parameters $\lambda_i$.

The relevant primary isotopic distribution is related to the observed 
distribution via a non-equilibrium evolution, which is described in simplest 
approximation by reaction kinetics where unstable nuclei feed the observed 
yields of stable nuclei.  As detailed in Ref.~\cite{roepke20}, taking into 
account all bound states below the edge of continuum states, the primary 
distribution  $Y^{{\rm rel},\gamma}_{A,Z}$ can be obtained, with 
Lagrange parameters $\lambda^\gamma_i$ obtained from a least squares fit to 
the observed final yields of $^2$H, $^3$H, $^4$He, $^6$He, and $^8$He.  The 
correct treatment of continuum states gives the virial expansion which is 
exact in the low-density limit.  Using measured scattering phase shifts, 
virial expansions have been determined for $^2$H, $^4$H, $^5$He and $^8$Be  
(which feeds $^4$He), see Ref.~\cite{horowitz06,roepke20_1}.  For the other 
isotopes estimates are given in~\cite{roepke20}.  Such a treatment including 
the continuum  states leads to a significant reduction in the calculated 
yields of the unbound nuclei $^4$He, $^5$He, $^7$He, and $^9$He.  As was 
shown in the ternary spontaneous fission of $^{252}$Cf~\cite{roepke20}, the 
observed yield of $^6$He is overestimated, and the observed yield of $^8$He 
is underestimated.  A possible explanation may be in-medium corrections, in 
particular Pauli blocking.  $^6$He is only weakly bound (the edge of 
continuum states is at 0.975 MeV which is small compared even to 2.225 MeV 
for $^2$H) so that Pauli blocking may dissolve the bound state at increasing 
density. To reproduce the observed yields, we have determined an effective 
pre-factor $R^{\rm rel, eff}_{A,Z}$. Both, the effective pre-factor and the 
relevant primary yields required to reproduce the observed yields are shown in 
Table \ref{Tab:3}. In detail, the pre-factor $R^{\rm rel, eff}_{A,Z}$ which 
represents the internal partition function was taken from the virial expansion 
for $^2$H, $^3$H, $^3$He, $^4$He, $^5$He, $^8$Be, as well as the estimates 
for $^8$He and $^9$He.  The corresponding observed yields are used to 
determine the three Lagrange parameters $\lambda_T$, $\lambda_n$, $\lambda_p$.
To reproduce the observed (weakly bound) $^6$He, the effective factor 
$R^{\rm rel,eff}_{^6{\rm He}}$ was determined.  For this, the contribution 
of the primary yields of $^6$He and $^7$He must be known. We used the 
value $Y_{^7{\rm He}}/Y_{^6{\rm He}} = 0.21$ measured for $^{252}$Cf 
in~\cite{kopatch02}.  It would be of interest to verify these predictions 
of $Y^{\rm rel,eff}_{A,Z}$ by measurements for $^{242}$Pu as were done 
for $^{252}$Cf.

Interpreting the effective pre-factors $R^{\rm rel, eff}_{A,Z}$ as reflecting 
in-medium corrections, we can use these inferred values to estimate the 
density.  These in-medium corrections are single-nucleon self-energy shifts 
which may be absorbed into the Lagrange parameters $\lambda_n$, $\lambda_p$ 
if momentum dependence of the single-particle self-energy is neglected. 
Then, the density dependence of $R^{\rm rel, eff}_{A,Z}$ is governed by the 
Pauli blocking effects which reduce the binding energies.  A global reduction 
of the binding energies is described in the generalized RMF approximation 
($x_{i,\sigma}=0.92$) given above by the effective cluster coupling to the 
mesonic field.  Within a more individual calculation, the Pauli blocking 
acts stronger for weakly bound states, eventually dissolving them, denoted 
as the Mott effect~\cite{roepke82,hagel12}.  We have performed an 
exploratory calculation assuming that Pauli blocking is essential for $^6$He 
because of its small binding but neglect the Pauli blocking shift for the 
stronger bound nuclei.  The reduction factor $R^{\rm rel,eff}_{^6{\rm He}}$ 
derived for $^6$He is smaller than the expected value 
$R^{\rm rel,vir}_{^6{\rm He}}=0.945$ according to the virial expansion. This 
leads to a shift of the binding energy of about 0.9 MeV and a correspondent 
density value of about $n_n = 0.0006$ fm$^{-3}$.  Note that this value has a 
large error because of uncertainties in the observed yield of $^6$He as well 
as the estimation of the energy shift of $^6$He due to in-medium corrections. 
Large deviations from the simple NSE are predicted for the primary yields of 
$^5$He, and it would be of interest to observe it like in the case of 
$^{252}$Cf~\cite{roepke20}.

A paper in which this approach is followed more consistently, considering also 
the Pauli blocking shifts for strongly-bound nuclei, and calculating 
the ternary fission yields for the $Z= 3-14$ isotopes observed 
in $^{241}$Pu($n_{\rm th}$,f)~\cite{koester00,koester99} is currently in 
preparation~\cite{roepke20_2}. 

 
\begin{table}[t]
\begin{tabular}{|c|c|c|c|c|c|}
\hline
isotope              & $Y_{A,Z}^{\rm exp}$ & $\frac{B_{A,Z}}{A}$ [MeV] & $g_{A,Z}$ &$R_{A,Z}^{\rm rel, eff}$ &$Y_{A,Z}^{\rm rel,eff}$        \\
\hline
$\lambda_T$ & -         & -                   & - & -        &   1.2042     \\
$\lambda_n$           & -         & -                   & - & -        & - 2.9954     \\
$\lambda_p$                     & -         & -                   & - & -        & -16.633      \\
\hline
$^1$n                  & -         & 0                   &2  &-         &1588200        \\
$^1$H                  & -         & 0                   &2  &-         &     19.16     \\
$^2$H                  & 42        & 1.112               &3  &0.98      &     42        \\
$^3$H$^{\rm obs}$                 & 786       & 2.827               &2  &-         &    786        \\
$^3$H                  & -         &2.827                &2  &0.99      &    779.51    \\
$^4$H                  & -         &1.720                &5  &0.0606    &      6.46679  \\
$^3$He                 & -         &2.573                &2  &0.988     &      0.004972 \\
$^4$He$^{\rm obs}$                & 10000     &7.073                &1  &-         &  10000        \\
$^4$He                 &-          &7.073                &1  &1         &   8485.89     \\
$^5$He                 &-          &5.512                &4  &0.7028   &   1508.81     \\
$^6$He$^{\rm obs}$                & 260       &4.878                &1  &-         &    260        \\
$^6$He                 &-          &4.878                &1  &0.8827  &     14.868    \\
$^7$He                 &-          &4.123                &4  &0.6235  &     45.122   \\
$^8$He$^{\rm obs}$                & 15        &3.925                &1  &-         &     15        \\
$^8$He                 &-          &3.925                &1  &0.9783    &     14.72     \\
$^9$He                 &-          &3.349                &2  &0.2604    &      0.27     \\
$^8$Be-                &-          &7.062                &1  &1.07      &      2.65     \\
\hline

\end{tabular}
\caption{Properties and relative yields of the H, He and Be isotopes from 
ternary fission $^{241}$Pu($n_{\rm th}$,f) which are relevant for the 
observed yields of H, He nuclei, (denoted by the superscript 'obs'.)   
Experimental  yields $Y^{\rm exp}_{A,Z}$~\cite{koester99} are compared to 
the yields calculated as described in the text.  Observed yields - column 2, 
binding energy $B_{A,Z}/A$ - column 3,  ground state degeneracy - column 4. 
prefactor - column 5, calculated primary isotope distribution - column 6.
}
\label{Tab:3}
\end{table}

\section{Summary and discussion}

In conclusion, experimentally determined reaction quotients have been 
determined for equatorially ejected isotopes of $Z \leq 14$ observed in the 
ternary fission of $^{242}$Pu.  The emission is characterized by $T =1.4$ MeV, 
$Y_p = 0.34$ and $\rho =2.6 \times10^{-4}$ nucleons/fm$^3$.  It should be 
noted that since at equilibrium the reaction coefficients are primarily 
sensitive to temperature and to density through medium effects extraction of 
accurate densities remains a difficult problem.  Here we have used the 
observed free neutron to free proton ratio to establish the density.

A comparison of the reaction quotients with those calculated using the EOS 
model of Pais \textit{et al.}~\cite{pais18} with a scaler cluster-meson 
coupling constant of $x_{i,\sigma}=0.92$ indicates a reasonable agreement 
between the experimental results and the model calculations for the lighter 
isotopes, indicating that chemical equilibrium is achieved for those isotopes 
and that medium effects are quite small at this temperature and density.  A 
more detailed evaluation of possible medium effects at these densities, 
addressing the properties of individual isotopes, is presented in 
reference~\cite{roepke20}.  The experimental yield of $^8$He is much higher 
than predicted in the calculation.  Other very neutron rich isotopes, 
$^9$Li, $^{10}$Be and $^{12}$Be, also give evidence of being underestimated 
in the calculation.  Whether this reflects the particular structural 
characteristics of these exotic nuclei warrants careful 
investigation~\cite{kisamori16,ivanytskyi19}.  For the heavier isotopes, the 
ratio of the measured reaction coefficient to the theoretically predicted 
equilibrium constant exponentially decreases with increasing mass.  This is 
attributed to a dynamical limitation, reflecting insufficient time for full 
equilibrium to develop~\cite{wuenschel14}.  An important point to be 
emphasized is that valid comparisons of calculated equilibrium constants to 
those derived from experimental data demand that the actual experimental 
ensemble of competing species be replicated as fully as possible in the 
calculation. 

\begin{table}[t]
\renewcommand{\thetable}{\Roman{table}(a)}
\begin{tabular}{|c|c|c|c|}
\hline
isotope & $Q_c^{\rm eff}$  expt. &  $Q_c^{\rm eff}$ calc. &  upper limit \\
\hline
$^2$H   & 6.61E+03  & 2.42E+04  &    \\
$^3$H   & 3.39E+09  & 3.30E+09  &    \\
$^4$H   & 3.63E+18  & 4.74E+18  &    \\
$^6$He  & 7.12E+25  & 5.46E+25  &    \\
$^8$He  & 3.09E+33  & 3.76E+32  &    \\
$^7$Li  & 1.54E+32  & 2.52E+32  &    \\
$^8$Li  & 2.66E+36  & 1.80E+36  &    \\
$^9$Li  & 1.44E+41  & 3.18E+40  &    \\
$^{11}$Li & 5.88E+46  & 7.87E+46  &    \\
$^7$Be  & 1.41E+34  & 2.33E+32  &**  \\
$^9$Be  & 2.34E+44  & 2.76E+44  &    \\
$^{10}$Be & 6.72E+49  & 2.20E+49  &    \\
$^{11}$Be & 2.37E+53  & 1.46E+53  &    \\
$^{12}$Be & 3.08E+57  & 1.14E+57  &    \\
$^{14}$Be & 2.24E+63  & 1.68E+63  &    \\
$^{10}$B  & 1.34E+50  & 1.08E+50  &**  \\
$^{11}$B  & 1.97E+56  & 2.65E+56  &    \\
$^{12}$B  & 3.37E+60  & 6.49E+60  &    \\
$^{14}$B  & 3.30E+68  & 1.34E+69  &    \\
$^{15}$B  & 3.21E+72  & 4.22E+72  &    \\
$^{17}$B  & 5.26E+79  & 2.05E+79  &**  \\
$^{14}$C  & 9.82E+73  & 1.65E+74  &    \\
$^{15}$C  & 9.20E+77  & 2.70E+78  &    \\
$^{16}$C  & 2.94E+82  & 4.87E+82  &    \\
$^{17}$C  & 1.03E+86  & 4.58E+86  &    \\
$^{18}$C  & 1.24E+90  & 3.19E+90  &    \\
$^{19}$C  & 3.04E+92  & 9.05E+93  &    \\
$^{20}$C  & 1.20E+97  & 4.77E+97  &    \\
\hline
\end{tabular}
\caption{ Experimental [26,27] and calculated equilibrium constants for 
light isotopes observed in the ternary fission of $^{242}$Pu. Assigned upper 
experimental limits are indicated  by **. See text for details. 
}
\label{Tab:A1}
\end{table}

\begin{table}[t]
\addtocounter{table}{-1}
\renewcommand{\thetable}{\Roman{table}(b)}
\begin{tabular}{|c|c|c|c|}
\hline
isotope & $Q_c^{\rm eff}$ expt. &  $Q_c^{\rm eff}$ calc. &  upper limit \\
\hline
$^{15}$N  & 2.89E+79  & 8.24E+80  &**  \\
$^{16}$N  & 1.42E+84  & 3.26E+85  &    \\
$^{17}$N  & 1.68E+89  & 1.92E+90  &    \\
$^{18}$N  & 2.17E+93  & 2.74E+94  &    \\
$^{19}$N  & 9.68E+97  & 2.56E+98  &    \\
$^{20}$N  & 2.96E+100 & 1.22E+103 &    \\
$^{21}$N  & 7.01E+104 &1.77E+107  &    \\
$^{15}$O  & 2.41E+83  & 5.06E+79  &**  \\
$^{19}$O  & 2.97E+101 & 8.11E+102 &    \\
$^{20}$O  & 3.45E+106 & 4.22E+107 &    \\
$^{21}$O  & 1.98E+110 & 3.20E+112 &    \\
$^{22}$O  & 2.83E+114 & 1.14E+117 &    \\
$^{24}$O  & 1.42E+123 & 1.77E+125 &**  \\
$^{19}$F  & 7.00E+103 & 1.37E+104 &    \\
$^{20}$F  & 1.92E+107 & 4.74E+109 &**  \\
$^{21}$F  & 5.54E+112 & 1.09E+115 &    \\
$^{22}$F  & 2.17E+118 & 1.31E+120 &    \\
$^{24}$F  & 1.64E+126 & 1.88E+129 &    \\
$^{24}$Ne & 6.69E+129 & 7.84E+132 &    \\
$^{27}$Ne & 2.77E+142 & 8.46E+145 &**  \\
$^{24}$Na & 5.13E+131 & 3.36E+134 &**  \\
$^{27}$Na & 5.30E+145 & 4.68E+149 &**  \\
$^{28}$Na & 1.75E+150 & 1.14E+154 &    \\
$^{30}$Na & 1.75E+158 & 3.94E+162 &**  \\
$^{27}$Mg & 3.90E+148 & 2.19E+152 &**  \\
$^{28}$Mg & 1.61E+154 & 4.45E+157 &**  \\
$^{30}$Mg & 6.05E+162 & 1.66E+167 &    \\
$^{30}$Al & 4.12E+165 & 5.20E+169 &**  \\
$^{34}$Si & 3.94E+186 & 1.53E+192 &**  \\
$^{35}$Si & 2.95E+191 & 6.67E+196 &**  \\
$^{36}$Si & 1.08E+196 & 2.12E+201 &**  \\
\hline
\end{tabular}
\caption{Continued: Experimental [26,27] and calculated equilibrium constants 
for light isotopes observed in the ternary fission of $^{242}$Pu. Assigned 
upper limits are indicated  by **. See text for details. 
}
\label{Tab:A2}
\end{table}

\section{Acknowledgements}
This work was supported by the United States Department of Energy under 
Grant \# DE-FG03-93ER40773, by the German Research Foundation (DFG), 
Grant \# RO905/38-1, the FCT (Portugal) Projects No. UID/FIS/04564/2019 and 
UID/FIS/04564/2020, and POCI-01-0145-FEDER-029912, and by PHAROS COST Action
CA16214. H. P. acknowledges the grant CEECIND/03092/2017 (FCT, Portugal).

\section{Appendix A} 

Tables \ref{Tab:A1} and \ref{Tab:A2} contain the $Q_c^{\rm eff}$ values 
presented in Figure 1 of this paper. We note that $Q_c$ values far above the 
calculated values are derived for the isotopes $^7$Be (parameter value 31) and 
$^{15}$O (parameter value 71).  Both values are based upon assigned upper 
limits. This comparison suggests that the actual yields for those two 
isotopes are well below these assigned values for upper limits. 




\begin{thebibliography}{}
\bibitem{oertel17}M. Oertel \textit{et al.}, Rev. Mod. Phys. \textbf{89}, 
015007 (2017).
\bibitem{lattimer91} J. M. Lattimer and F. Douglas Swesty, 
Nucl. Phys. A \textbf{535}, 331 (1991). 
\bibitem{shen98} H. Shen, H. Toki, K. Oyamatsu, K. Sumiyoshi, 
Nucl. Phys. \textbf{A637}, 435 (1998); 
Prog. Theor. Phys. \textbf{100}, 1013 (1998).
\bibitem{horowitz06}C. J. Horowitz and A. Schwenk, 
Phys. Lett. \textbf{B638}, 153 (2006); Nucl. Phys. A 776, 55 (2006).
\bibitem{typel10} S. Typel et al., Phys. Rev. C \textbf{81}, 015803 (2010).
\bibitem{hempel10}M. Hempel and J. Schaffner-Bielich, 
Nucl. Phys. \textbf{A837}, 210 (2010).
\bibitem{shen11} H. Shen, H. Toki, K. Oyamatsu, and K. Sumiyoshi, 
Astrophys. J. Suppl. \textbf{197}, 20 (2011).
\bibitem{shen11_1} G. Shen, C. J. Horowitz, and E. O'Connor, 
Phys. Rev. C \textbf{83}, 065808 (2011).
\bibitem{voskresenskaya12}M. D. Voskresenskaya and S. Typel, 
Nucl. Phys. \textbf{A887}, 42 (2012).
\bibitem{roepke13}G. R\"opke \textit{et al.}, 
Nucl. Phys. \textbf{A897}, 70 (2013).
\bibitem{furusawa13}S. Furusawa \textit{et al.}, 
Astrophys. J. \textbf{772}, 95 (2013). 
\bibitem{fsu} B.  G.  Todd-Rutel  and  J.  Piekarewicz,  
Phys.  Rev.  Lett. {\bf 95}, 122501 (2005).
\bibitem{pais18}H. Pais \textit{et al.}, Phys. Rev. C \textbf{97}, 045805 (2018).
\bibitem{pais19}H. Pais \textit{et al.}, 
Phys. Rev. Lett. \textbf{125}, 012701 (2020).
\bibitem{chomaz14}Dynamics and Thermodynamics~with Nuclear~Degrees of 
Freedom, Euro. Phys. J. \textbf{30}, Issue 1. (2006) P. Chomaz \textit{et al.} 
editors.
\bibitem{bali14}''Topical issue on Nuclear Symmetry Energy.'' 
Eur. Phys. J. A \textbf{50} (2014).
Bao-An Li \textit{et al.}, editors.
\bibitem{elliot12}J. Elliott \textit{et al.}, arXiv:1203.5132 (2012).
\bibitem{qin12}L. Qin \textit{et al.},  
Phys. Rev. Lett. \textbf{108}, 172701(2012).
\bibitem{hempel15}Matthias Hempel \textit{et al.}, 
Phys. Rev. C \textbf{91}, 045805 ( 2015).
\bibitem{wagemans91}C. Wagemans, The Nuclear Fission Process (CRC Press, 
Boca Raton, 1991). and references therein.
\bibitem{mills95}J. Mills, Fission Product Yield Evaluation, 
Thesis Univ. of Birmingham (1995). 
\bibitem{iaea00}IAEA-TECDOC-1168 ISSN 1011–4289 IAEA, Vienna, (2000). 
and references therein.
\bibitem{halpern71}I. Halpern, Ann. Rev. Nucl. Sci \textbf{21}, 245 (1971).
\bibitem{theobald89}J. P. Theobald, Nucl. Phys. \textbf{A502} 343 (1989).
\bibitem{mehta73}G. K. Mehta \textit{et al.}, 
Phys. Rev. C \textbf{7}, 373 (1973).
\bibitem{serot98}O. Serot and C. Wagemans, Nucl. Phys. \textbf{A641}, 34 (1998).
\bibitem{heeg89}P. Heeg \textit{et al.}, 
Nucl.  Inst. \&   Meth. \textbf{A278}, 452 (1989).
\bibitem{koester00}U. Koester Dissertation, Technischen 
Universit\"at M\"unchen (2000).
\bibitem{koester99}U. Koester \textit{et al.}, 
Nucl. Phys. \textbf{A652}, 371 (1999).
\bibitem{lestone05}J. P. Lestone LA-UR-05-8860 (2006), 
Phys. Rev. C \textbf{72}, 014604 (2005).
\bibitem{tsekhanovich}I. Tsekhanovich, \textit{et al.}, 
Phys. Rev. C \textbf{67}, 034610 (2003).
\bibitem{shafer95}R. Schafer and T. Fliessbach, 
J. Phys. G \textbf{21}, 861 (1995).
\bibitem{rubchenya88}V. Rubchenya and S.Z. Yavshits, Z. Phys. \textbf{A329}, 217 (1988).
\bibitem{rubchenya82}V. A. Rubchenya Yad. Fiz. \textbf{35}, 576 (1982). 
\bibitem{kopatch02}Yu. N. Kopatch \textit{et al.}, 
Phys. Rev. C \textbf{65}, 044614 (2002).
\bibitem{vorobyev18}A. S. Vorobyev \textit{et al.}, 
J. Expt. And Theor. Phys. \textbf{127}, 659 (2018).
\bibitem{goennenwein05}F. Goennenwein, M. Mutterer and Y. Kopatch, Europhys.
 News, \textbf{36}, 11 (2005).
\bibitem{vermote10}S. Vermote \textit{et al.}, Seminar on Fission :  
Het Pand, Gent, Belgium, 17-20 May 2010 /  editors, Cyriel Wagemans, 
Jan Wagemans, Pierre D'hondt. World Scientific Page 145.
\bibitem{piasecki75}E. Piasecki \textit{et al.}, 
Nucl. Phys. \textbf{A255}, 387 (1975).
\bibitem{piasecki79}E. Piasecki and L. Nowicki, Proc. Symp. Phys. and 
Chem. Of Fission (Juelich 1979) IAEA, Vienna (1979) p 193.
\bibitem{sharma83}M. Sharma, PhD Thesis, Indian Institute of Technology, 
(1983).
\bibitem{ramayya07}A. V. Ramayya, J.H. Hamilton and J.K. Hwang, 
Rom. Rept. Phys. \textbf{59}, 595 (2007).
\bibitem{schubert92}A. Schubert, J. Hutsch and  K. M\"uller, 
Z. Phys. A - Hadrons and Nuclei \textbf{341}, 481-488 (1992).
\bibitem{wagemans08}C. Wagemans \textit{et al.}, 
Phys. Rev. C \textbf{78}, 064616 ( 2008 ).
Europhys. News. Jan./Feb. (2005).
\bibitem{wuenschel14}S. Wuenschel \textit{et al.}, 
Phys. Rev. C \textbf{90}, 011601(R) (2014).
\bibitem{roepke20}G. R\"opke, J. B. Natowitz and H. Pais, submitted to 
Eur. J. Phys. A, August  2020.
\bibitem{demo97}P. Demo and Z. Kozisek, 
J. Phys. G \textbf{23}, 971 (1997).
\bibitem{schmelzer97}J. Schmelzer and G. R\"opke, 
Phys. Rev. C \textbf{55}, 
1917 (1997).
\bibitem{wilemski95}G. Wilemski and B.E. Wyslouzil, 
J. Chem. Phys. \textbf{103}, 1127 (1995).
\bibitem{mekjian78}A. Z. Mekjian, Phys. Rev. C \textbf{17}, 1051 (1978).
\bibitem{hagel00}K. Hagel \textit{et al.}, 
Phys. Rev. C \textbf{62}, 034607 (2000). 
\bibitem{andronenko01}M. Andronenko, \textit{et al.}, 
Eur. Phys. Jour. A \textbf{12}, 185 (2001).
\bibitem{popkiewicz96}M. Popkiewicz \textit{et al.} (1996),\\ https://www.osti.gov/etdeweb/servlets/purl/583067.  
\bibitem{sobotka97}L. G. Sobotka \textit{et al.}, 
Phys. Rev. C \textbf{55}, (1997).
\bibitem{hagel12}K. Hagel \textit{et al.}, 
Phys. Rev. Lett. \textbf{108}, 062702 (2012).
\bibitem{cervesato92}I. Cervesato \textit{et al.},
Phys. Rev. C \textbf{45}, 2369 (1992).
\bibitem{ensdf} BNL ENSDFENSDF  Evaluated Nuclear Structure Data File. 
https://www.nndc.bnl.gov/].
\bibitem{roepke82}G. R\"opke, L. M\"unchow and H. Schulz, 
Nucl. Phys. \textbf{A379}, 536 (1982).
\bibitem{oertzen11}W. von Oertzen, Int. 
J. Mod. Phys. E, \textbf{20}, 765 (2011).
\bibitem{kisamori16}K. Kisamori, \textit{et al.}, 
Phys. Rev. Lett. \textbf{116}, 052501 (2016).
\bibitem{ivanytskyi19}O. Ivanytskyi, M. Angeles Perez-Garcia, and C. Albertus, 
Eur. Phys. J. A \textbf{55} 184 (2019).
\bibitem{roepke20_1}G. R\"opke, Phys. Rev. C \textbf{101}, 064310 (2020).
\bibitem{roepke20_2}G. R\"opke \textit{et. al.} paper in preparation.

\end{thebibliography}
\end{document}